\documentclass[%
 amsmath,amssymb,
 aps,
]{revtex4-2} 

\usepackage{amsmath,amsthm}
\usepackage{graphicx}
\usepackage{dcolumn}
\usepackage{bm}
\usepackage{color}
\usepackage{epsfig}
\usepackage{multirow}
\usepackage{mathrsfs}
\usepackage{hyperref}
\usepackage{cleveref}
\usepackage{epstopdf}
\usepackage{makecell}
\usepackage[linesnumbered,ruled]{algorithm2e}
\usepackage{subfigure}
\usepackage{booktabs}
\usepackage{tabularx}

\begin{document}
\preprint{Preprint}

\title{Pythagoras Superposition Principle for Localized Eigenstates of 2D Moir\'e Lattices}

\author{Zixuan Gao}
\affiliation{School of Mathematical Sciences, CMA-Shanghai, MOE-LSC and Shanghai Center for Applied Mathematics, Shanghai Jiao Tong University, Shanghai 200240, China}

\author{Zhenli Xu}
\affiliation{School of Mathematical Sciences, CMA-Shanghai, MOE-LSC and Shanghai Center for Applied Mathematics, Shanghai Jiao Tong University, Shanghai 200240, China}

\author{Zhiguo Yang}
\affiliation{School of Mathematical Sciences, CMA-Shanghai, MOE-LSC and Shanghai Center for Applied Mathematics, Shanghai Jiao Tong University, Shanghai 200240, China}

\author{Fangwei Ye}
\affiliation{School of Physics and Astronomy, Shanghai Jiao Tong University, Shanghai, 200240, P. R. China}



\begin{abstract}
	
Moir\'e lattices are aperiodic systems formed by a superposition of two periodic lattices with a relative rotational angle.
In optics, the photonic moir\'e lattice has many appealing properties such as its ability to localize light, thus attracting much attention on exploring features of such a structure. One fundamental research area for photonic moir\'e lattices is the properties of eigenstates, particularly the existence of localized eigenstates and the localization-to-delocalization transition in the energy band structure. Here we propose an accurate algorithm for the eigenproblems of aperiodic systems by combining plane wave discretization and spectral indicator validation under the higher-dimensional projection, allowing us to explore energy bands of fully aperiodic systems. A localization-delocalization transition regarding the intensity of the aperiodic potential is observed and a novel Pythagoras superposition principle for localized eigenstates of 2D moir\'e lattices is revealed by analyzing the relationship between the aperiodic and its corresponding periodic eigenstates. This principle sheds light on exploring the physics of localizations for moir\'e lattice.

\end{abstract}

\maketitle

\section{Introduction}
The structural geometrical properties of natural or artificial systems profoundly impact the properties of waves that are allowed to propagate in them. Thus, a fascinating range of phenomena stemming from the geometrical properties of material landscapes, such as their periodicity, are continuously discovered in diverse areas of physics, including mechanics, acoustics, optics, electronics, solid-state physics, and physics of matter waves \cite{bistritzer2011moire,cao2018unconventional,carr2017twistronics,gu2022dipolar,lau2022reproducibility,han2001moire,gonzalez2019cold,o2016moire,hu2020moire,lu2019superconductors}. {Recently, moir\'e systems \cite{lou2021theory,sharpe2019emergent,fan2002analysis,dos2007graphene,sinha2022berry,cao2018correlated} have drawn much attention due to their unusual electronic, optical and magnetic properties, and their potential for designing novel materials with tailored functionalities \cite{andrei2021marvels,du2023moire}. A moir\'e system is a system that involves two or more periodic structures with different lattice constants or orientation, which interact with each other to form a spatial moir\'e pattern.} These systems can arise in a variety of fields such as materials science, condensed matter physics, optics, and electronics. For example, in condensed-matter physics, the moir\'e systems are revealing a wealth of profound physical effects that have established a new area of research referred to as twistronics \cite{carr2017twistronics}. Moir\'e patterns are also crucial in all areas of physics related to wave propagation, such as Bose-Einstein condensates and optics \cite{o2016moire,hu2020moire}, where they afford the possibility to explore the phenomena that arise because of the transition from aperiodic (incommensurate) to periodic (commensurate) geometries, occurring at specific values of the rotation angle in contrast to aperiodic quasicrystal systems.

Photonic moir\'e lattices can be created by the superposition of two rotated square or hexagonal sublattices \cite{wang2020localization,huang2016localization}.
Recent experiments reported the observation of the 2D localization-delocalization transition (LDT) \cite{wang2020localization} of light waves when one tunes the twisting angles or the depth of the modulation of the constitute sublattices.
In one dimension the LDT effect had been observed for both light \cite{lahini2009observation} and matter waves \cite{billy2008direct}. The localization phenomenon is due to the band flattening of the moir\'e pattern in the incommensurate (namely, aperiodic) phase \cite{wang2020localization}. Theoretically, the properties of localized eigenstates and the existence of the LDT in the eigenvalue spectrum for a fixed moir\'{e} lattice remain less explored due to the difficulty in calculating eigenproblems for aperiodic systems. Traditional crystalline approximant methods \cite{goldman1993quasicrystals,lifshitz1997theoretical} are slowly convergent and cannot provide results accurate enough for physical understanding. In this Article, we developed an efficient method to solve the aperiodic problems, which allows us to explore the properties of localized eigenstates in moir\'e lattices, and by this method we reveal the Pythagoras superposition principle between aperiodic system and its periodic crystalline approximants.

In the paraxial approximation, the propagation of an extraordinarily polarized beam in a photorefractive medium
with an optically induced refractive index is described by the Schr\"{o}dinger-like equation of the dimensionless field amplitude $\phi(\bm{r},z)$ \cite{efremidis2002discrete}:
\begin{equation}\label{001}
\mathrm{i}\dfrac{\partial \phi}{\partial z}=-\dfrac{1}{2}\nabla^2\phi+\dfrac{E_0}{1+I(\bm{r})}\phi
\end{equation}
where $\bm{r}=(x,y)$ and $I(\bm{r})=|p_1v(\bm{r})+p_2v(S\bm{r})|^2$ is the intensity of the moir\'e lattice induced by two ordinarily polarized mutually coherent periodic sublattices, $v(\bm{r})$ and $v(S\bm{r})$. Here $S$ is a 2D rotational matrix such that $S(\theta)\bm{r}$ rotates vector $\bm{r}$ by a counterclockwise angle $\theta$.
$p_1$ and $p_2$ are the amplitudes of the first and second sublattices, and $p_1/p_2$ is defined as the lattice ratio.
The potential term $V(\bm{r})=E_0/(1+I(\bm{r}))$ describes the optical response of the photorefractive crystal \cite{wang2020localization,huang2016localization}{, where $E_0$ describes the strength. Notably, the moir\'e structures can be periodic or aperiodic, depending on the twisting angle $\theta$.  The moir\'e lattices composed of two square lattices as considered here, they are periodic when $\theta$ takes any Pythagorean angle and aperiodic otherwise \cite{huang2016localization}.} Throughout this article, the sublattices of the moir\'e systems have fixed parameter $p_1/p_2=1$.

\begin{figure*}[htbp]
	\centering
	\includegraphics[width=1.0\textwidth]{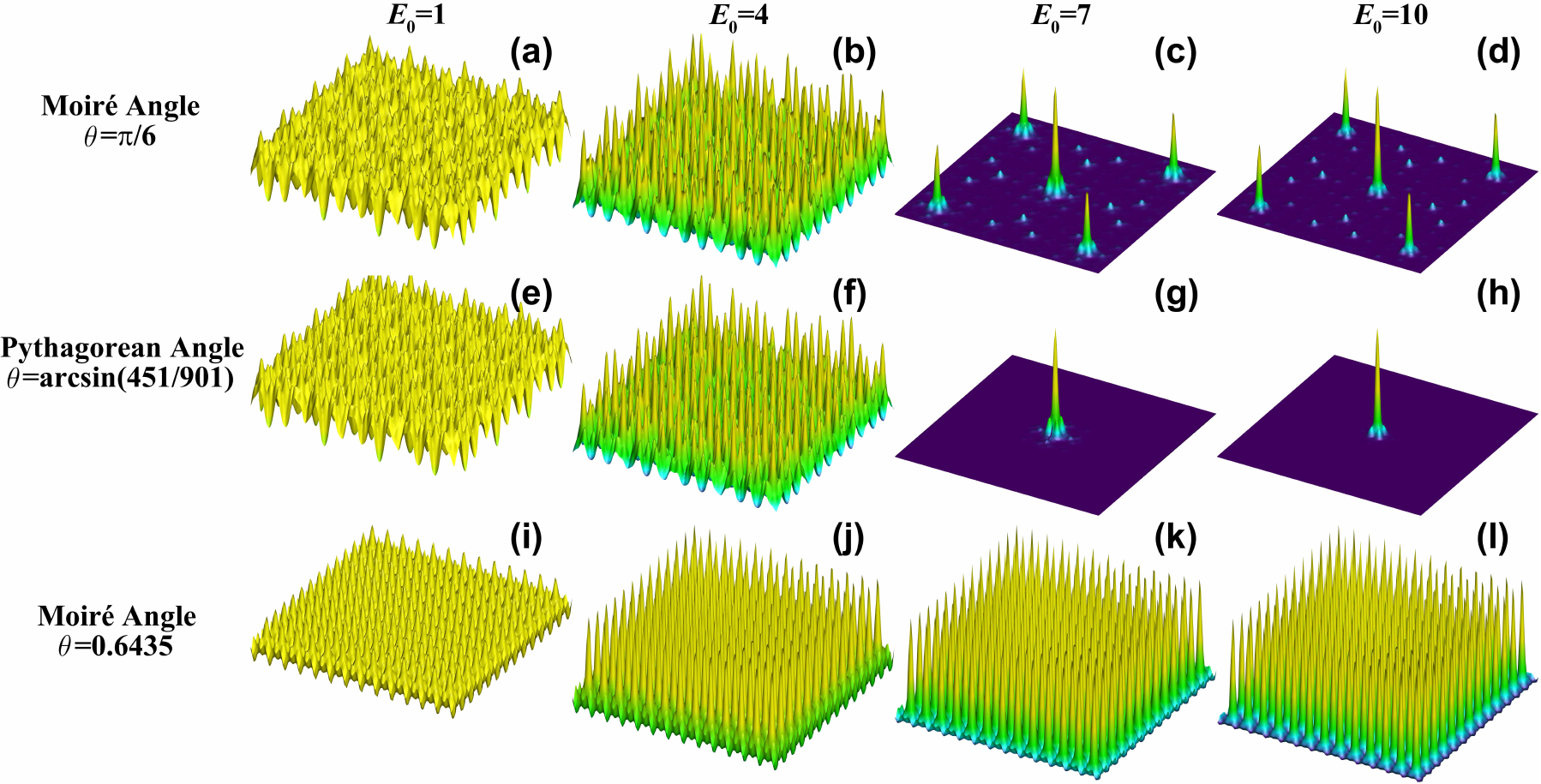}
\caption{The first eigenstates $|\psi|^2$ of the lattices for $E_0=1,4,7$ and $10$ with twist angles (a-d)$\theta=\pi/6$; (e-h) $\theta=\arcsin(451/901)$; and (i-l) $\theta=0.6435$.
Here the $\theta=\arcsin(451/901)$ cases correspond to a periodic system whose twist angle approximates $\pi/6$.}
 \label{fig1}
\end{figure*}

To visualize the mysterious properties of moir\'e lattices, Fig. \ref{fig1} presents the first eigenstates of aperiodic systems for
different $E_0$ with $\theta=\pi/6$ and periodic crystalline approximants
with Pythagorean angle $\theta=\arcsin(451/901)$ which approximates $\pi/6$ with error $\sim 2\times 10^{-4}\pi$,
together with the results of aperiodic systems with $\theta=0.6435$ which approximates the Pythagorean angle $\arcsin(3/5)$. These data describes the results in domain $[-33.3401, 33.3401]^2$. Here the
aperiodic systems are calculated by the projection indicator (PI) method described below, while the periodic systems are solved by the plane wave method \cite{shen2011spectral}.
The photonic lattice actually acts as an effective potential that can trap or diffuse light during its propagation and the larger modulus indicates a stronger effective potential.
These results clearly show a more localized tendency with the increase of $E_0$. At $E_0=7$ and $10$ both the aperiodic
and periodic systems demonstrate mode localization. We note that, although the lattice for (e-h) is periodic, the width of the wavepacket in $E_0=7,\theta=\arcsin(451/901)$ is 2.813,
which is much less than the period $T=\sqrt{901/2}\pi$ implying that the eigenstate of the periodic approximant is localized. This localized eigenstate shows that a localized light field having the central ring shape in the transverse plane is obtained, indicating a localized high-order light mode \cite{zeng2021localization}. Oppositely, for $\theta=0.6435$ which is close to the Pythagorean angle $\arcsin(3/5)$, the eigenstates are delocalized since it can be approximated by a periodic system with much smaller period $T=\sqrt{5/2}\pi$.

Interestingly, the eigenstate of the aperiodic systems features a lot of sharp peaks and looks completely different from that of its periodic companion,
comparing between $\theta=\pi/6$ and $\arcsin(451/901)$. Different from the non-localized light fields, peaks of the localized moir\'{e} light field are more like a superposition of several ring-shape localized light fields rather than side lobes around the central main lobes indicating a weakly diffusing effect \cite{zeng2021localization}. Each peak corresponds to a localized light field in the neighborhood and the photonic moir\'{e} lattices have some ability to trap the light during its propagation. This is counterintuitive, taking into account that the difference in rotation angles of two systems
is negligible, illustrating the discontinuity in eigenstates with the rotation angle. Thus, generally, this tells the breakdown of the traditional crystalline approximant method \cite{goldman1993quasicrystals,lifshitz1997theoretical}
to solve the eigenstates of the aperiodic systems and the error cannot be controlled due to the simultaneous Diophantine approximation \cite{davenport1946simultaneous}. {Another commonly employed approach for addressing similar problems is the continuum model Hamiltonian method \cite{bistritzer2011moire}. This method solves the Hamiltonian structure \cite{dos2007graphene} of the moir\'{e} system in the specific area. Instead of truncating the traditional domain in physic space, the continuum model Hamiltonian method truncates the momentum-space lattice at the first shell, thereby rendering a truncation error of momentum space similar to the conventional crystalline approximant method. This motivates us to propose an efficient and accurate method for aperiodic eigenvalue problems that circumvents the error like simultaneous Diophantine error.}

\section{Method}

Here we develop the projection indicator (PI) method to solve the eigenproblem of the Schr\"odinger equation
\begin{equation}\label{1dsys}
E\psi=-\dfrac{1}{2}\nabla^2\psi+\dfrac{E_0}{1+I(\bm{r})}\psi,
\end{equation}
 by a combination of the projection method \cite{jiang2014numerical} and the indicator projection with plane wave
discretizations. {For simplicity, one chooses the projection matrix $\bm{P}$ so that the period in each direction is $2\pi$, expressed as
\begin{equation}
   \bm{P}=2\begin{bmatrix}
    \cos{\gamma} & -\sin{\gamma} & \cos{\gamma} & \sin{\gamma}\\
     \sin{\gamma} & \cos{\gamma} & -\sin{\gamma} & \cos{\gamma}
    \end{bmatrix}
\end{equation}
with $\gamma=\theta/2$, where the projection direction is in parallel to the shortest periodic edge of the 4D periodic system. By the projection matrix $\bm{P}$, one can obtain the transition from 2D to 4D spaces, $\bm{q}=\bm{P}^\top\bm{r}$, with $\bm{q}=(q_1,q_2,q_3,q_4)$.} Then substituting $\phi=e^{-iEt}\psi$ into Eq. \eqref{001} leads to a 4D eigenproblem,
\begin{equation}\label{003}
E\psi=-\dfrac{1}{2}\sum_{i,j=1}^4\dfrac{\partial^2 \psi}{\partial q_i\partial q_j}\left(\dfrac{\partial q_i}{\partial x}\dfrac{\partial q_j}{\partial x}+\dfrac{\partial q_i}{\partial y}\dfrac{\partial q_j}{\partial y}\right)+\tilde{V}(\bm{q})\psi.
\end{equation}
{Here $\tilde{V}(\bm{q})$ represents the potential function $V(\bm{r})$ after being lifted to 4D space, which implies that if $\bm{q}=\bm{P}^\top\bm{r}$ holds, then $\tilde{V}(\bm{q})=V(\bm{r})$. $\tilde{V}(\bm{q})$ is a periodic function with period $[0,2\pi]^4$, thus Eq. \eqref{003} is a 4D periodic eigenproblem. The numerical solution $\tilde{\psi}_N$ of Eq. \eqref{003} can be expanded using the plane wave expansion as
\begin{equation}\label{spex}
\tilde{\psi}_N(\bm{q})=\sum_{\bm{k}\in\Omega}\psi_{\bm{k}}e^{\mathrm{i}\langle \bm{k},\bm{q}\rangle}.
\end{equation}
Here $\langle\cdot,\cdot\rangle$ denotes the standard inner product between vectors, $\Omega=\mathbb{Z}^{4}\cap\{||\bm{k}||_{\infty}\leq N\}$ is  the basis space, and $N$ represents the number of spectral modes in each dimension, and $\psi_{\bm{k}}$ are the Fourier expansion coefficients. Let $\bm{k}=(k_1,\cdots,k_{4})$. Eq. \eqref{003} can be transformed into
\begin{equation}\label{coreeq}
    E\psi_{\bm{k}}=\dfrac{1}{2}\sum_{i=1}^2\sum_{j=1}^{4}\sum_{l=1}^{4}\left(\psi_{\bm{k}}k_jk_l\dfrac{\partial q_j}{\partial r_i}\dfrac{\partial q_l}{\partial r_i}\right)+\mathcal{F}\{V\psi\}_{\bm{k}},
\end{equation}
where $\mathcal{F}\{\cdot\}$ denotes the Fourier transform and $\mathcal{F}\{V\psi\}_{\bm{k}}$ is the Fourier coefficient of $\mathcal{F}\{V\psi\}$ with the frequency $\bm{k}$. Set a column vector $\vec{\psi}$ containing all the Fourier expansion coefficients $\psi_{\bm{k}}$. Hence, one can form a matrix $\bm{A}$ to transform Eq. \eqref{coreeq} into a matrix eigenproblem $\bm{A}\vec{\psi}=E\vec{\psi}$. Due to the enormous size of $\bm{A}$ which is not sparse, it cannot be stored explicitly. Therefore, we use a matrix-free preconditioned Krylov subspace method \cite{liesen2013krylov} which only requires the matrix-vector product to be stored in each iteration, making it a more efficient approach. Once the eigenvector $\vec{\psi}$ is obtained, the four-dimensional eigenfunction $\psi(\bm{q})$ can be approximated using Eq. \eqref{spex}. By the projection matrix $\bm{P}$, $\tilde{\psi}_N$ can be transformed back to the 2D space, which implies that
\begin{equation}
\tilde{\psi}_N(\bm{q})=\sum_{\bm{k}\in\Omega}\psi_{\bm{k}}e^{\mathrm{i}\langle\bm{k},\bm{q}\rangle} 
=\sum_{\bm{k}\in\Omega}\psi_{\bm{k}}e^{\mathrm{i}\langle \bm{Pk},\bm{r}\rangle}
:={\psi}_N(\bm{r}).
\end{equation}
Here the second equality uses $\langle \bm{k},\bm{P}^{\top}\bm{r}\rangle=\langle \bm{Pk},\bm{r}\rangle$.
The 2D aperiodic function ${\psi}_N(\bm{r})$ as the eigenfunction and $E$ as the eigenvalue are the numerical results of the original problem Eq. \eqref{001} returned by the PI. Therefore, this approach is equivalent to solving the higher-dimensional periodic lattice problem within the subspace of the original problem, applying the plane wave method. On the other hand, if the twisting angle is such that the moir\'e lattices restore periodicity, one can solve the corresponding eigenproblem directly in the original (2D) space, using the plane wave method, thanks to the Floquet-Bloch theorem \cite{joannopoulos2008molding}.}

Due to the singularity of the eigenvalue problem, especially when $E_0$ is large, the error in the numerical calculation may lead to some spurious eigenstates. In order to sift out pseudo-eigenstates, a spectral indicator method \cite{liu2019spectral,huang2020multilevel,kato2013perturbation} is adopted.
Define the indicator \cite{huang2016recursive}
\begin{equation}
    \mathrm{Ind}=\|\bm{Q} (\bm{Qf}/\|\bm{Qf}\|)\| 
\end{equation}
where matrix $\bm{Q}$ is the spectral projection 
\begin{equation} \label{integral}
\bm{Q}=\dfrac{1}{2\pi i}\int_\Gamma(\bm{A}-s\bm{I})^{-1}ds,
\end{equation}
with $\bm{I}$ being the identity matrix. The indicator becomes 1 if there exists at least one eigenvalue in the square. To evaluate the integral, the closed path is divided uniformly into four parts and the composite  trapezoidal rule is employed to compute $\bm{Qf}$ where $\bm{f}$ takes the potential function $V$ directly. {The numerical approximation of $\bm{Qf}$ can be obtained via a certain quadrature rule
\begin{equation}
    \bm{Qf}\approx\dfrac{1}{2\pi i}\sum_{j=1}^{n_0}\omega_j \bm{r}_j.
\end{equation}
Here $\{\omega_j\}$ are quadrature weights and $\{\bm{r}_j\}$ are the solutions of the linear systems
\begin{equation}\label{gmres}
    (\bm{A}-s_j \bm{I})\bm{r}_j=\bm{f},j=1,2,\dots,n_0,
\end{equation}
where $\{s_j\}$ are the quadrature nodes on $\Gamma$. For simplicity, the piecewise trapezoid formula is chosen for the curve integral Eq. \eqref{integral}.} The size of $\Gamma$ is small such that only a few sample points guarantee high accuracy. Since the spectral projection method provides many solutions of Eq. \eqref{003}, we can take a small domain around each eigenvalue and calculate the indicator value to validate the correctness of this eigenvalue.  Detailed steps of the PI are included in Algorithm \ref{algorithm1}.

\begin{algorithm}\label{algorithm1}
\KwData{$d$-dimensional quasi-periodic potential $V$, the number of bases $N$ in each dimension, the number of eigenvalues $M$, the step size $\delta$, and the threshold value $\epsilon$}

Determine the basis space and the test space $\Omega$ 

Compute the first $M$ eigenpairs of $\bm{A}$ as $\{(E_m,u_m)\}_{m=1}^M$ 

\For{$m=1$ to $M$}
{Set $\omega=[E_m-\delta/2,E_m+\delta/2]^2$ and $\bm{f}$ by $V$

Compute the indicator $\mathrm{Ind}=\|\bm{Q} (\bm{Qf}/\|\bm{Qf}\|)\|$ by Eq. \eqref{integral}

\If{$\mathrm{Ind}<\epsilon$}
{Delete the eigenpair $(E_m,u_m)$}

Project the eigenfunctions $u_m$ back into the $d$-dimensional space}

\caption{Projection indicator method}
\end{algorithm}

\section{Accuracy performance}

\begin{figure}[htbp]
	\centering
	\includegraphics[width=0.5\textwidth]{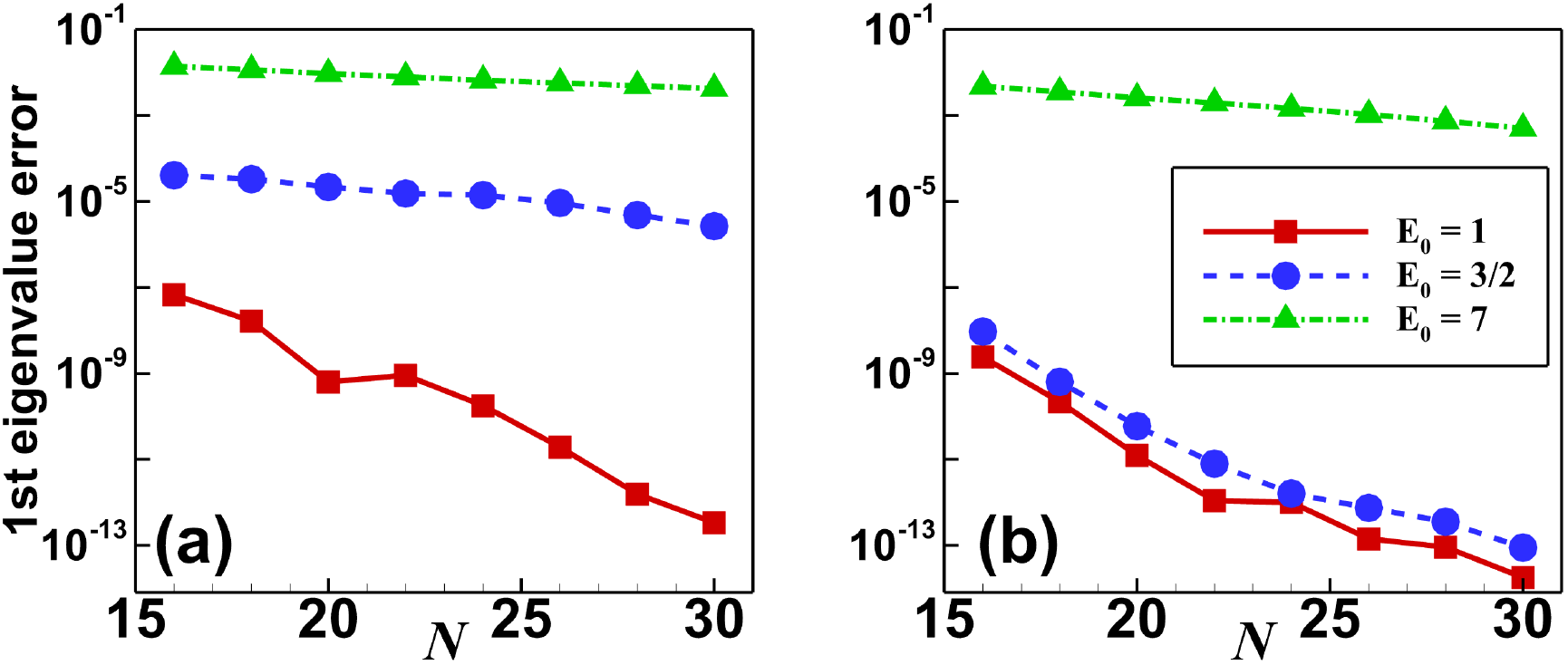}
	\caption{Error of the first eigenvalue of moir\'e systems with the increase of $N$ for $E_0=1, 3/2$ and $7$: (a) the 1D case; (b) the 2D case.}\label{appfig55}
\end{figure}

{To validate the accuracy performance of the PI method, the absolute errors of the first eigenvalues  are evaluated and presented  in Fig. \ref{appfig55} where panels  (a) and (b) correspond to 1D and 2D moir\'e systems, respectively. Here the 1D moir\'e system is with potential $V(x)=E_0/(1+I(x))$ for $I(x)=(\cos(2x\cos\theta)+\cos(2x\sin\theta))^2+1$, and the rotation angle is set as $\pi/6$. The results for three cases of $E_0=1,3/2$ and $7$ are presented with the increase of $N$. In these low strength cases $E_0=1,3/2$, the eigenstates are delocalized and one can see the spectral convergence of the numerical approximation. Conversely, in the strong strength case $E_0=7$, the numerical approximation converges slower and more nodes are needed to achieve high accuracy. These results demonstrate that the larger $E_0$, the more singular the system, and the slower the convergence of numerical results. In 1D, a small value of $N$ can achieve an accuracy of approximately $10^{-10}$ for $E_0=1$. For 2D systems, rapid convergences are observed for both cases of $E_0=1$ and $E_0=3/2$. These results demonstrate the high accuracyand the fast convergence  of the PI method for low strength cases.}

\begin{figure}[htbp]
	\centering
	\includegraphics[width=0.5\textwidth]{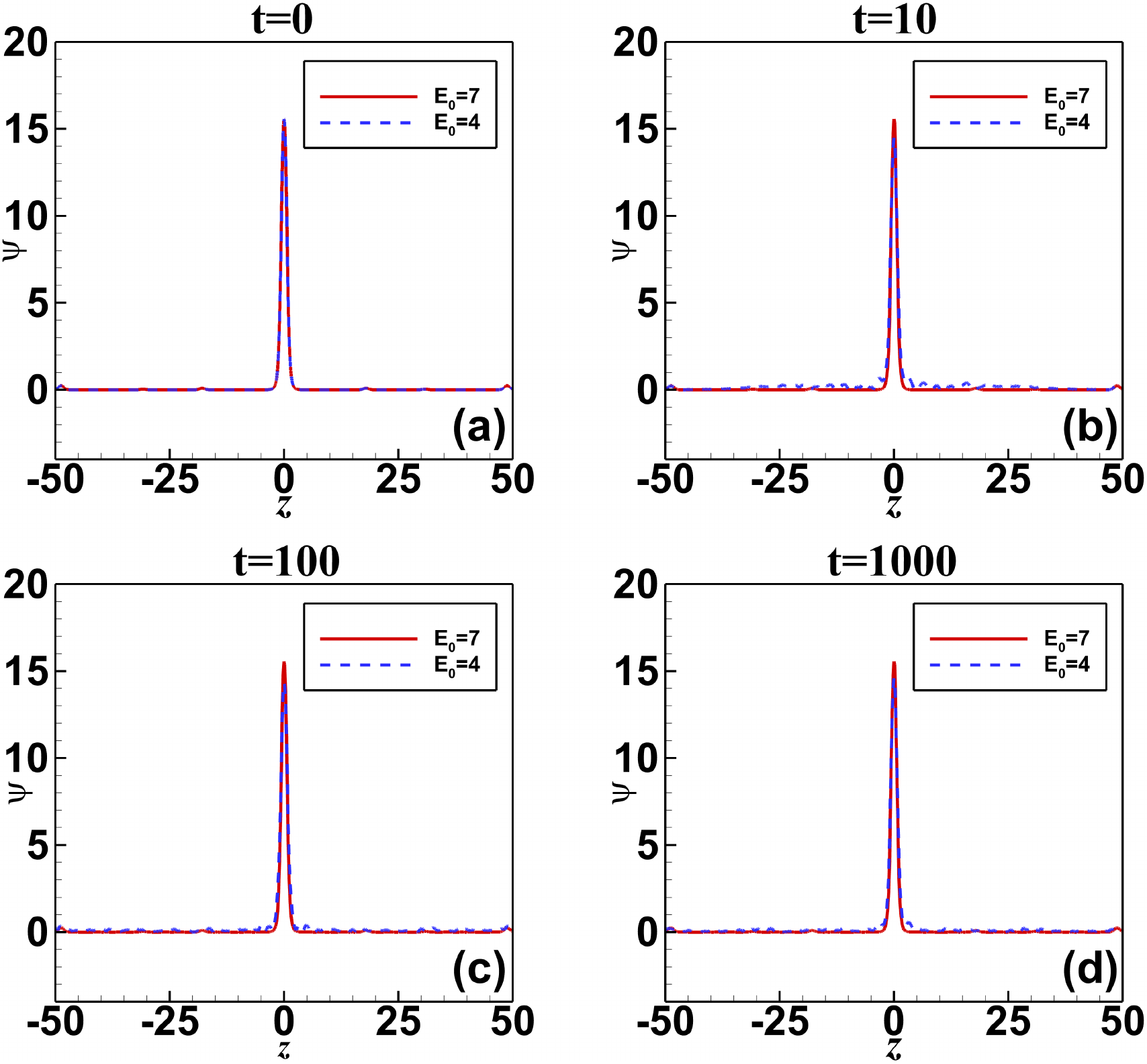}
	\caption{Wave propagation of 1D aperiodic potentials at $E_0=4$ and $7$. (a,b,c,d) correspond ot the results t $t=0,10,100$ and $1000$. }\label{appfig1-2}
\end{figure}

{For systems with large strength $E_0$, the eigenfunctions tend to be localized. This makes numerical calculation more challenging to achieve high accuracy. 
Consequently, if the numerical error in the eigenfunction is significant, the wave function of the time evolution will be unable to remain unchanged 
during the propagation. By examining the wave propagation behavior by time-dependent Schr\"odinger equation Eq. \eqref{001}, one can validate the accuracy of the eigenstates obtained through the PI. Fig. \ref{appfig1-2} presents the results of the 1D moir\'e system at different times. The initial condition takes the eigenfunctions for $E_0=4$ and 7, which is calculated using the PI with $N=50$. One can observe that for the case of $E_0=7$ 
the wave function remains unaltered regardless of the propagation time. However, for a smaller strength $E_0=4$, the wave function exhibits a slight oscillation
changes during propagation. The behavior shows the stable propogation of the localized eigenstate and the results demonstrate that the PI method provides
accurate solution even when the eigenfunctions are localized.}

\begin{figure}[h]
	\centering
	\includegraphics[width=0.5\textwidth]{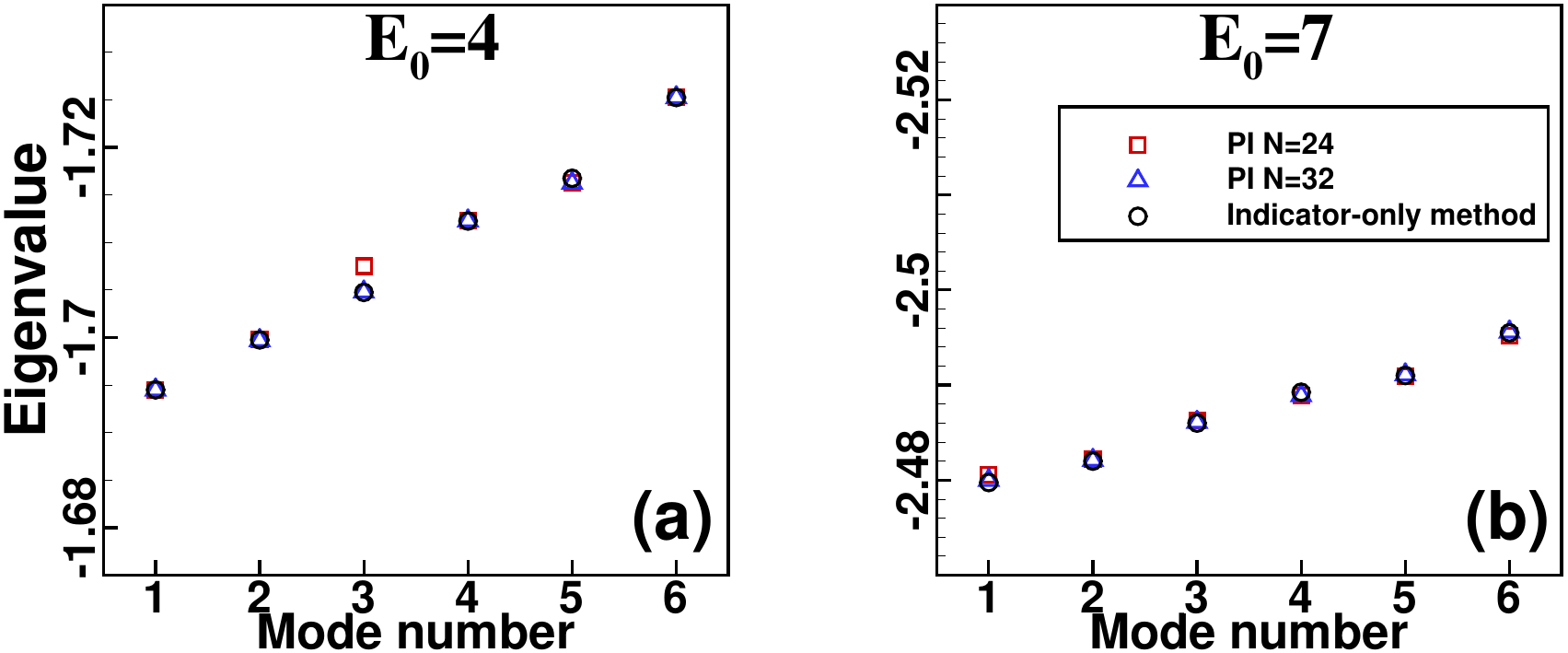}
	\caption{The eigenvalues obtained by the PI and the indicator-only method. The first six eigenvalues of 2D aperiodic systems with $E_0=4$ and $7$ are displayed.
The projection method uses plane waves $N=24$ and $32$ along one direction. The indicator method uses intervals of size $2.5\times10^{-4}$.}\label{appfig2}
\end{figure}

{To further validate the accuracy of the PI for localized eigenstates, we calculate the first 6 eigenvalues of the 2D aperiodic systems with $E_0=4$ and 7 and display the results in Fig. \ref{appfig2}.  In the spectral projection step, 24 and 32 spectral nodes are used in each dimension, and the spectral indicator sets the size of $\Gamma$ to be $2.5\times10^{-4}$. Notably, the overall errors are less than $10^{-3}$, serving as evidence of the high accuracy of the PI. The indicator-only method is used to verify the accuracy of the PI results, which involves searching for eigenvalues across the entire domain. For the results obtained through the spectral projection step, one finds that less than half of the eigenvalues are validated by the indicator method. This highlights the necessity of the indicator test in ensuring accuracy.}

\section{Pythagoras Superposition Principle}
\begin{figure*}
	\centering
	\includegraphics[width=1.0\textwidth]{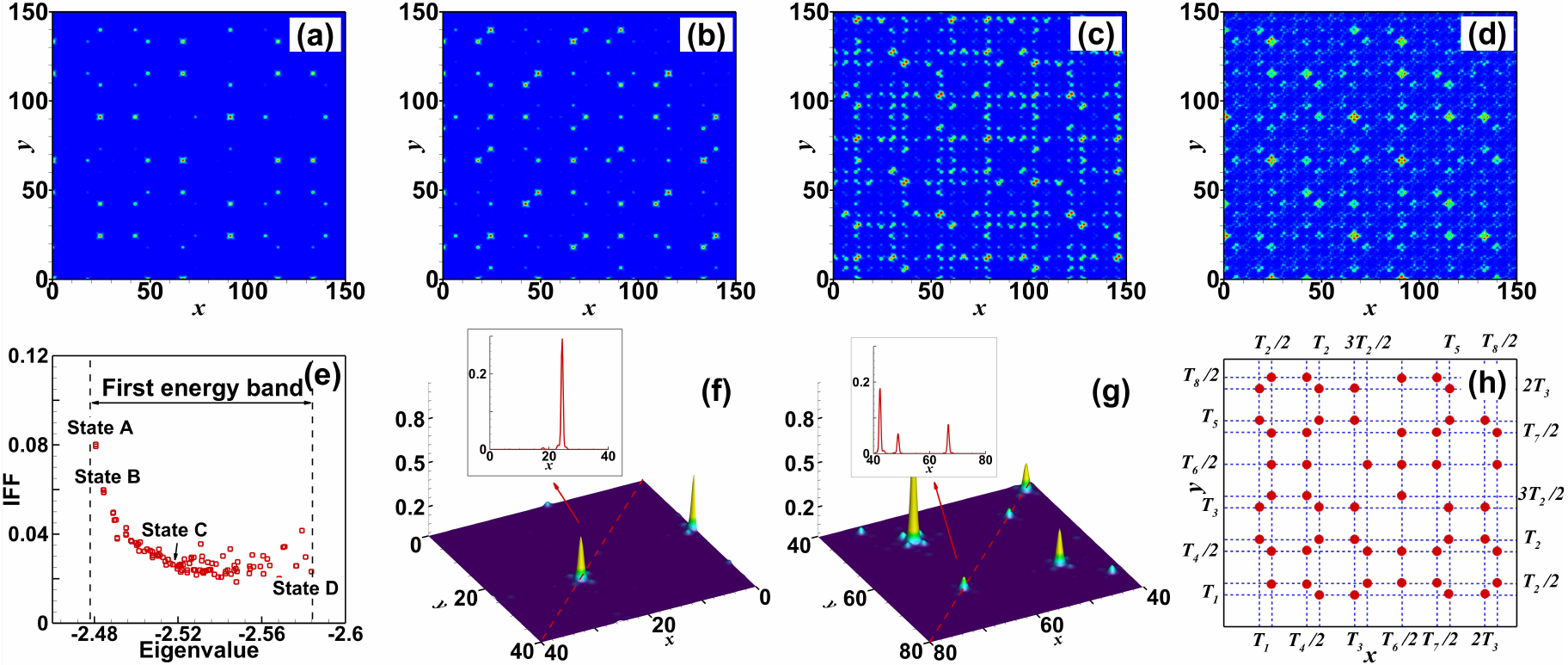}
	\caption{Results of the 2D aperiodic system at $E_0=7$ and $\theta=\pi/6$. (a,b) Contours of the first and third eigenstates; (c,d) Contours of the eigenstates at the middle ($41^{\mathrm{st}}$) and bottom ($109^{\mathrm{th}}$) of the first energy band;
(e) The IFFs of all the eigenstates in the first energy band as a function of eigenvalue. States A-D correspond to the eigenstates of panels (a-d);
(f,g) Enlarged plots of the first eigenstate in two domains $[0,40]^2$ and $[40,80]^2$ with the $y=x$ cuts; (h) The peak sites of interior wavepackets (red circles)  in panel (a), which are located at the nodes of the $T$ mesh.}\label{fig3}
\end{figure*}

We are now ready to study the energy bands and the LDT phenomenon of 2D photonic moir\'e lattices. The LDT in moir\'e lattice with respect to the change of the lattice ratio was discussed in Wang {\it et al.} \cite{wang2020localization}. However, the effect of varying $E_0$ on localization remains unclear. {The PI is used to calculate the eigenstates of the moir\'e system. Through the location of the peaks in the localized eigenstates, one can deduce the localization position of the wavepackets propagated in this system.} Here we consider the 2D aperiodic systems at $E_0=7$ and $\theta=\pi/6$ and attempt to understand the misconvergence of the simultaneous Diophantine approximation.
Fig. \ref{fig3}(a-d) show the $1^{\mathrm{st}}, 3^{\mathrm{rd}}, 41^{\mathrm{st}}$ and $109^{\mathrm{th}}$ eigenstates for $|\psi|^2$, with $N=30$. The two eigenstates of (c,d) are at the middle and bottom of the first energy band of the aperiodic system,
where the degree of localization decreases with the increase of mode numbers. A quantitative description of the
localization degree of the eigenstates in a given region $U$ is the integral form factor (IFF) \cite{wang2020localization,schwartz2007transport} expressed by,
\begin{equation}
\mathrm{IFF}=\dfrac{\left(\int_U |\psi|^4 d^2\bm{r}\right)^{1/2}}{\int_U |\psi|^2 d^2\bm{r}}.
\end{equation}
 A larger IFF means a more localized state of eigenfunction $\psi$. Fig. \ref{fig3}(e) displays the IFFs of all eigenstates in the first energy band, where
states A-D correspond to panels (a-d), respectively. One observes the decreasing tendency of the IFF value with the more index.

The first eigenstate of the aperiodic system is the only one with an IFF bigger than 0.05. In order to verify the localization character,
the enlarged plots of the eigenstate are present in Fig. \ref{fig3}(f,g), together with the $y=x$ cuts for $|\psi|^2$. The exponential decay of the wavepackets can be observed, demonstrating the exponential localization characteristics of the eigenfunction. Fig. \ref{fig3}(a-d) thus verifies the mode transition from localization into delocalization with the increasing of the mode index in the energy spectrum. This is in agreement with the experimental demonstration \cite{wang2020localization}
of LDT in 2D photonic moir\'e lattices, which revealed the mechanics for wave localization based on flat-band structure, in contrast to the schemes based on light diffusion in photonic quasicrystals requiring disorder media \cite{freedman2006wave,levi2011disorder}.

\begin{figure}[htbp]
	\centering

        \includegraphics[width=0.5\textwidth]{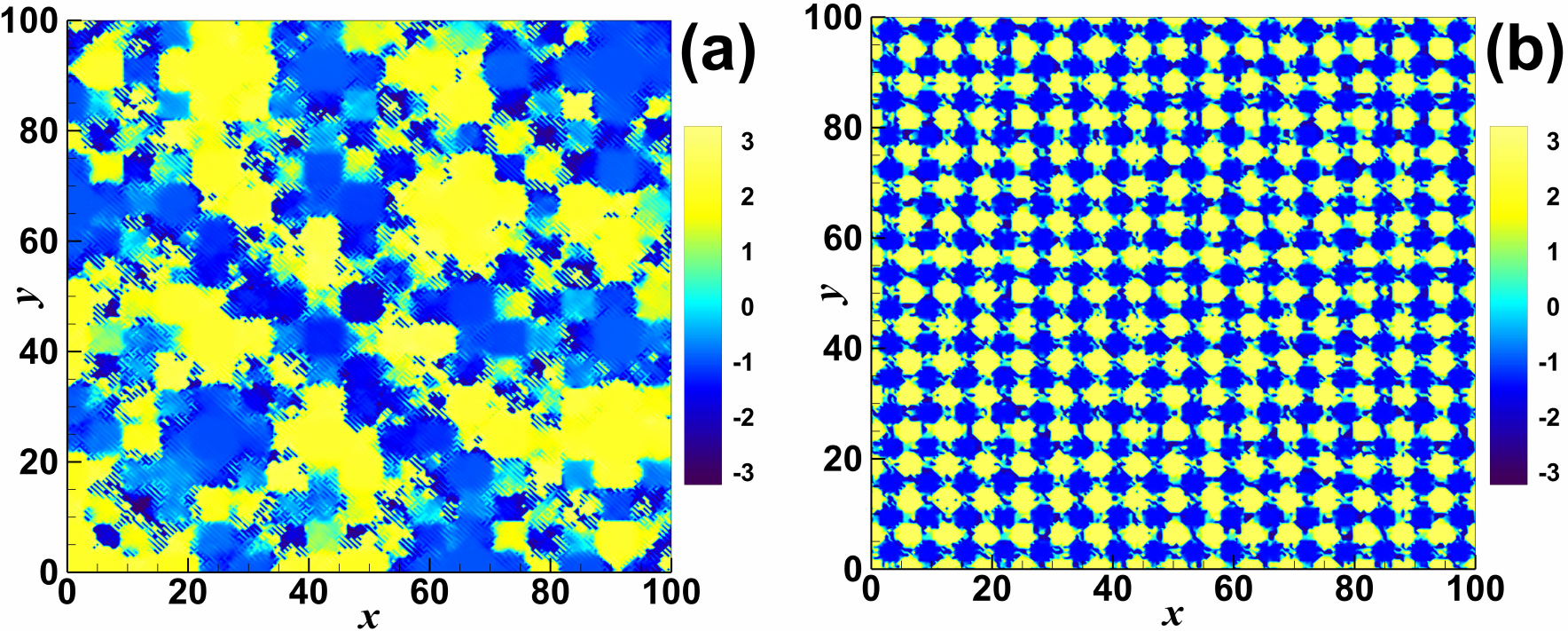}
        \caption{The phase structure of the first eigenstate of PI: (a) the aperiodic localized system, (b) the periodic approximant.}\label{appfig8}
\end{figure}

{The difference between aperiodic eigenstates and their periodic approximants can be also illustrated by the phase structures of the first eigenstates. Fig. \ref{appfig8} displays the results of the localized system with parameters  $E_0=7$ and $\theta=\pi/6$ and its periodic approximant. For localized systems, as depicted in Fig. \ref{appfig8} (a),  the phase does not change very frequently with space. In some areas, there also exist abrupt changes in phase, which implies that the moir\'{e} lattice can also preserve the topological phase structure of the localized light mode. While in Fig. \ref{appfig8} (b), the phase of periodic approximants changes periodically and strongly. These differences show that direct periodic approximation may not preserve all physical properties.}


Moreover, Fig. \ref{fig1} indicates that the crystalline approximants do not converge to the aperiodic system. Consequently, the IFFs of the aperiodic system are significantly smaller than the results
of its periodic counterpart. In order to connect the relations between the moir\'e lattice and its crystalline approximants, we introduce
a Pythagoras triple $(a,b,c)$ to represent a Pythagoras angle
such that $\sin\theta=a/c.$ Let the central cell be $[0,T]^2$ where  $T=\sqrt{c}\pi$ or $\sqrt{c/2}\pi$ is the period. Due to the symmetry of the potential, the first eigenfunction is composed of wavepackets located at the four corners of the cell for period $\sqrt{c/2}\pi$, or of wavepackets at the corners and the center of the cell for period $\sqrt{c}\pi$. Table \ref{tb1} lists the $c$, $\theta$ and $T$ values of the 9 periodic approximants to the moir\'e angle (with period $T_i<150$, $i=1,\cdots, 9$).
Fig. \ref{fig3}(h) displays that the peak sites of those interior wavepackets are all located at the nodes of the $T$ mesh (grid points  $nT_i/2$ for integer $n$) for the first eigenstate shown in Fig. \ref{fig3}(a). One can clearly observe that these happen to be the packet sites by all these periodic approximants. This counterpart clearly appears as a superposition principle for localized eigenstates of the Pythagoras angles. This is to say that an eigenstate of the moir\'e lattice can be considered as the summation of the eigenstates of its crystalline approximants, and the
weight of each approximant depends on the Diophantine error between the twist angles. The periodic systems near $\theta=\pi/6$ have large periods, while the periodic system with the smallest period, $\theta=\arcsin(3/5)$, is near $\theta=0.6435$. By the superposition principle, when $E_0$ is large the eigenstates of the former systems show sparse peak distribution, while the peaks of the eigenstates of $\theta=0.6435$ are very dense. This conclusion is in consistent with the numerical results in Fig. \ref{fig1}.

\begin{table}
	\caption{\label{tb1}Parameters of 9 periodic systems near $\pi/6$}
		\begin{tabular}{p{0.15\textwidth}p{0.15\textwidth}p{0.15\textwidth}}
                \hline\hline
			$c$   & $\theta$    & $T$\\
			\midrule
			\specialrule{0em}{1pt}{1pt}
			$65$ & $0.169500\pi$ & $T_1=\sqrt{65/2}\pi$ \\
			\specialrule{0em}{1pt}{1pt}
			$241$ & $0.165903\pi$ & $T_2=\sqrt{241}\pi$ \\
			\specialrule{0em}{1pt}{1pt}
			$901$ & $0.166858\pi$ & $T_3=\sqrt{901/2}\pi$\\
			\specialrule{0em}{1pt}{1pt}
			$725$ & $0.167431\pi$ & $T_4=\sqrt{725}\pi$ \\
			\specialrule{0em}{1pt}{1pt}
			$2701$ & $0.166476\pi$ & $T_5=\sqrt{2701/2}\pi$ \\
			\specialrule{0em}{1pt}{1pt}
			$3361$ & $0.166603\pi$ & $T_6=\sqrt{3361}\pi$ \\
			\specialrule{0em}{1pt}{1pt}
			$4813$ & $0.167081\pi$ & $T_7=\sqrt{4813}\pi$ \\
			\specialrule{0em}{1pt}{1pt}
			$7925$ & $0.163487\pi$ & $T_8=\sqrt{7925}\pi$ \\
			\specialrule{0em}{1pt}{1pt}
			$10085$ & $0.166731\pi$ & $T_9=\sqrt{10085}\pi$ \\
                \hline\hline
		\end{tabular}
\end{table}

\begin{figure*}[htbp]
	\centering
	\includegraphics[width=0.75\textwidth]{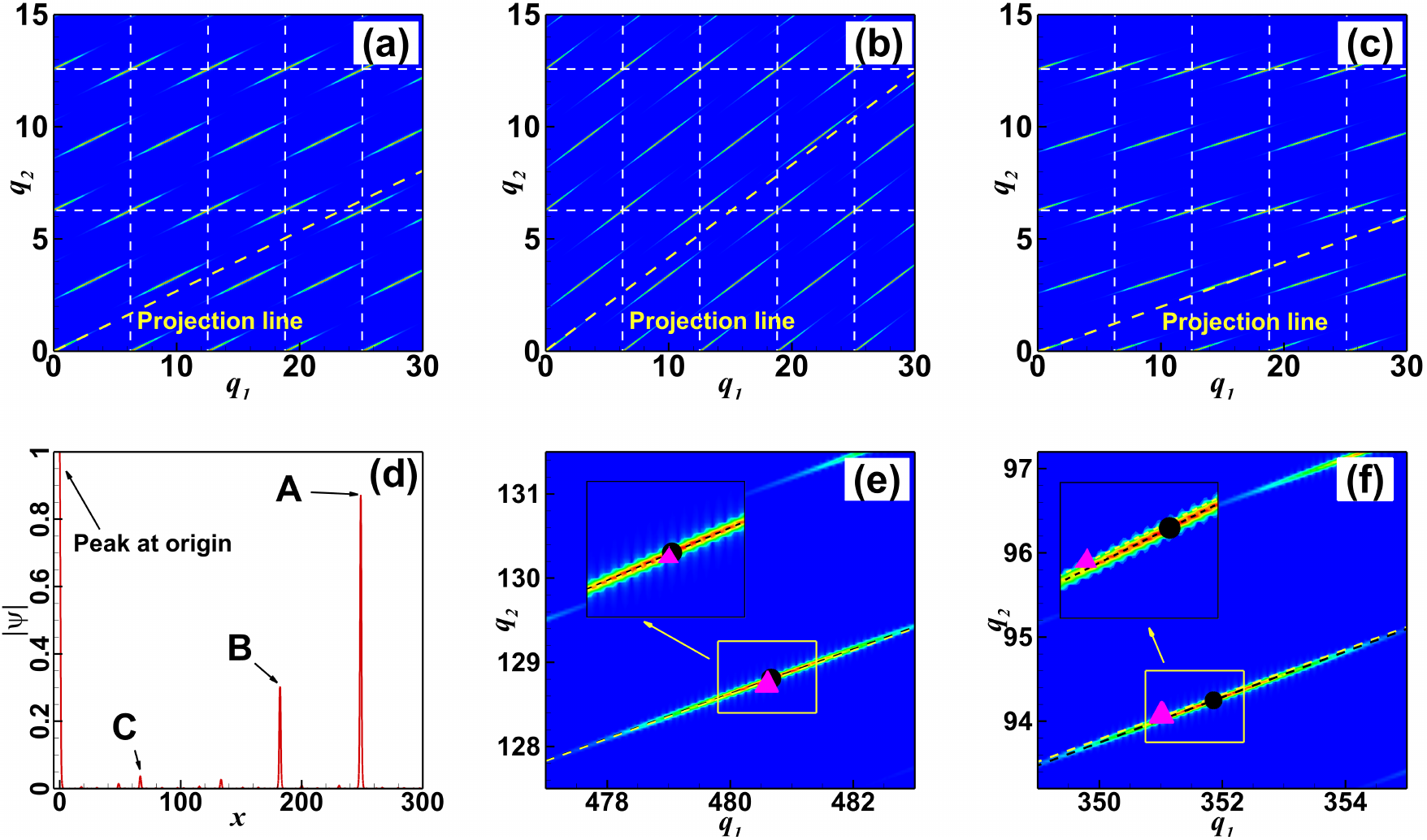}
	\caption{The first eigenstate of the 1D aperiodic system: (abc) the contour plots in the incremental space of the rotation angle $\pi/6,\pi/4$ and $\pi/8$. The corresponding projection lines are $q_2=\tan(\pi/12)q_1,q_2=\tan(\pi/4)q_1$ and $q_2=\tan(\pi/16)q_1$, respectively; (d) the projected eigenstate in the physical space and the widths of A-C are all about $3.68$;
(ef) the detailed contours corresponding to peaks A and B. The rotation angle of (def) is $\pi/6$.}\label{fig4}
\end{figure*}

To provide a theoretical understanding of the Pythagoras superposition principle, we consider the simplified 1D aperiodic system whose equation is Eq .\eqref{1dsys}. Here the use of the 1D system is due to its intuitive physical picture and the corresponding Schr\"odinger-like equation is also easy to solve in the incremental space $(q_1,q_2)$. Fig. \ref{fig4}(a-c) shows the contour of the first eigenstates of different rotation angles $\pi/6,\pi/4$ and $\pi/8$ in the $\bm{q}$ space, where some localized regions of javelin shape can be observed, and we denote them as the localization in higher dimensions. The projection lines defined by $q_2=\tan(\pi/12)q_1,q_2=\tan(\pi/4)q_1$ and $q_2=\tan(\pi/16)q_1$ describe the physical solutions of the aperiodic systems. {Since each projection line is not parallel to the javelin-shaped domain, it intersects with many domains, leading to the wavepackets of the physical space in a lower dimension. The localization phenomenon is independent of the rotation angle.} Fig. \ref{fig4}(d) shows these packets for $x<300$, where peaks A-C are due to the periodic systems with $c=12545,3361$ and $901$, correspondingly the periods are $\sqrt{12545/2}\pi,\sqrt{3361}\pi$ and $\sqrt{901}\pi$. These are in agreement with the analysis based on the Pythagoras triple. The periodic system due to peak A has rotation angle $\arcsin(6273/12545)\approx0.16668\pi$, an error of $10^{-5}\pi$ to the moir\'e angle $\pi/6$. This small Diophantine error results in a strongly localized wavepacket, as observed from the figure.

Fig. \ref{fig4}(c-d) illustrates the intersecting lines corresponding to peaks A and B in panel (b), where the black dash lines are
the projection lines of the periodic systems, slightly different from the projection line of the aperiodic system. In the enlarged subplots, the purple triangles represent the locations of these peaks, and the black circles are the corresponding peak locations of the periodic systems. Due to the periodic approximation, the black line crosses the central axis of the
javelin-shaped domain. One can observe that in each panel, the triangle and circle symbols are very close, demonstrating
that the locations of wavepackets for the aperiodic systems do have a relation to the approximate periodic systems.
Let $\varepsilon$ be the distance between the two symbols, representing the error in the locations of the wavepackets.
This error can be roughly estimated as  $\varepsilon\approx T\Delta\theta$, where $T$ is the period of the periodic system
and $\Delta\theta$ is the difference between the twist angles of the aperiodic and periodic systems. $\varepsilon$ values of peaks A-C are $0.0115,0.0332$ and $0.0427$, respectively. $T\Delta\theta$ values are $0.0115,0.0313$ and $0.0428$, which are high precision approximations of $\varepsilon$. Hence $T\Delta\theta$ can characterize the error in the localizations of the wavepackets.

\section{Conclusion}
To summarize, we propose a highly efficient PI algorithm, which is a combination of the  projection method and the indicator for
aperiodic eigenproblems for photonic moir\'e lattices. The PI solves the problem directly without using periodic approximations
such that the simultaneous Diophantine approximation can be avoided. It allows us to accurately calculate the band structure of the eigenstates
in moir\'e lattices. {In addition, we conduct analysis on connections between periodic and aperiodic systems in terms of their structures, numerical algorithms and eigenstate properties.} We find that the localized eigenstates in the moir\'e lattices are determined by the periodic lattices adjacent to it,
leading to the Pythagoras superposition principle. This principle connects the relationship between the aperiodic and periodic lattices
and is promising to further explore the moir\'{e} lattices and the wavepacket localization in 2D and 3D systems.


\begin{acknowledgements}
Z. G. and Z. X. are supported by the National Natural Science Foundation of China (NNSFC)(No. 12071288) and Science and Technology Commission of Shanghai Municipality (grant Nos. 20JC1414100 and 21JC1403700). Z. Y. is supported by the NNSFC (No. 12101399) and the Shanghai Sailing Program (No. 21YF1421000). F.Y. is supported by the NNSFC (No. 91950120), Scientific funding of Shanghai (No. 9ZR1424400), and Shanghai Outstanding Academic Leaders Plan (No. 20XD1402000).
\end{acknowledgements}



\begin{thebibliography}{35}
\expandafter\ifx\csname natexlab\endcsname\relax\def\natexlab#1{#1}\fi
\expandafter\ifx\csname bibnamefont\endcsname\relax
  \def\bibnamefont#1{#1}\fi
\expandafter\ifx\csname bibfnamefont\endcsname\relax
  \def\bibfnamefont#1{#1}\fi
\expandafter\ifx\csname citenamefont\endcsname\relax
  \def\citenamefont#1{#1}\fi
\expandafter\ifx\csname url\endcsname\relax
  \def\url#1{\texttt{#1}}\fi
\expandafter\ifx\csname urlprefix\endcsname\relax\def\urlprefix{URL }\fi
\providecommand{\bibinfo}[2]{#2}
\providecommand{\eprint}[2][]{\url{#2}}

\bibitem[{\citenamefont{Bistritzer and MacDonald}(2011)}]{bistritzer2011moire}
\bibinfo{author}{\bibfnamefont{R.}~\bibnamefont{Bistritzer}} \bibnamefont{and}
  \bibinfo{author}{\bibfnamefont{A.~H.} \bibnamefont{MacDonald}},
  \bibinfo{journal}{Proceedings of the National Academy of Sciences}
  \textbf{\bibinfo{volume}{108}}, \bibinfo{pages}{12233}
  (\bibinfo{year}{2011}).

\bibitem[{\citenamefont{Cao et~al.}(2018)\citenamefont{Cao, Fatemi, Fang,
  Watanabe, Taniguchi, Kaxiras, and Jarillo-Herrero}}]{cao2018unconventional}
\bibinfo{author}{\bibfnamefont{Y.}~\bibnamefont{Cao}},
  \bibinfo{author}{\bibfnamefont{V.}~\bibnamefont{Fatemi}},
  \bibinfo{author}{\bibfnamefont{S.}~\bibnamefont{Fang}},
  \bibinfo{author}{\bibfnamefont{K.}~\bibnamefont{Watanabe}},
  \bibinfo{author}{\bibfnamefont{T.}~\bibnamefont{Taniguchi}},
  \bibinfo{author}{\bibfnamefont{E.}~\bibnamefont{Kaxiras}}, \bibnamefont{and}
  \bibinfo{author}{\bibfnamefont{P.}~\bibnamefont{Jarillo-Herrero}},
  \bibinfo{journal}{Nature} \textbf{\bibinfo{volume}{556}}, \bibinfo{pages}{43}
  (\bibinfo{year}{2018}).


\bibitem[{\citenamefont{Carr et~al.}(2017)\citenamefont{Carr, Massatt, Fang,
  Cazeaux, Luskin, and Kaxiras}}]{carr2017twistronics}
\bibinfo{author}{\bibfnamefont{S.}~\bibnamefont{Carr}},
  \bibinfo{author}{\bibfnamefont{D.}~\bibnamefont{Massatt}},
  \bibinfo{author}{\bibfnamefont{S.}~\bibnamefont{Fang}},
  \bibinfo{author}{\bibfnamefont{P.}~\bibnamefont{Cazeaux}},
  \bibinfo{author}{\bibfnamefont{M.}~\bibnamefont{Luskin}}, \bibnamefont{and}
  \bibinfo{author}{\bibfnamefont{E.}~\bibnamefont{Kaxiras}},
  \bibinfo{journal}{Physical Review B} \textbf{\bibinfo{volume}{95}},
  \bibinfo{pages}{075420} (\bibinfo{year}{2017}).

\bibitem[{\citenamefont{Gu et~al.}(2022)\citenamefont{Gu, Jie, Ma, Liguo, Liu, Song, Watanabe, Kenji, Taniguchi, Takashi, Hone, James C and Shan, Jie and Mak, Kin Fai}}]{gu2022dipolar}
\bibinfo{author}{\bibfnamefont{J.}~\bibnamefont{Gu}},
  \bibinfo{author}{\bibfnamefont{L.}~\bibnamefont{Ma}},
  \bibinfo{author}{\bibfnamefont{S.}~\bibnamefont{Liu}},
  \bibinfo{author}{\bibfnamefont{K.}~\bibnamefont{Watanabe}},
  \bibinfo{author}{\bibfnamefont{T.}~\bibnamefont{Taniguchi}},
  \bibinfo{author}{\bibfnamefont{J.~C.}~\bibnamefont{Hone}},
  \bibinfo{author}{\bibfnamefont{J.}~\bibnamefont{Shan}},
  \bibnamefont{and}
  \bibinfo{author}{\bibfnamefont{K.}~\bibnamefont{Mak}},
  \bibinfo{journal}{Nature Physics} \textbf{\bibinfo{volume}{18}},
  \bibinfo{pages}{395} (\bibinfo{year}{2022}).

\bibitem[{\citenamefont{Lau et~al.}(2022)\citenamefont{Lau, Chun Ning and Bockrath, Marc W and Mak, Kin Fai and Zhang, Fan}}]{lau2022reproducibility}
\bibinfo{author}{\bibfnamefont{C.}~\bibnamefont{Lau}},
  \bibinfo{author}{\bibfnamefont{M.~W.}~\bibnamefont{Bockrath}},
  \bibinfo{author}{\bibfnamefont{M.}~\bibnamefont{Fai}},
  \bibnamefont{and}
  \bibinfo{author}{\bibfnamefont{Z.}~\bibnamefont{Fan}},
  \bibinfo{journal}{Nature} \textbf{\bibinfo{volume}{602}},
  \bibinfo{pages}{41} (\bibinfo{year}{2022}).

\bibitem[{\citenamefont{Han et~al.}(2001)\citenamefont{Han, B and Post, D and Ifju, P}}]{han2001moire}
\bibinfo{author}{\bibfnamefont{B.}~\bibnamefont{Han}},
  \bibinfo{author}{\bibfnamefont{D.}~\bibnamefont{Post}},
  \bibnamefont{and}
  \bibinfo{author}{\bibfnamefont{P.}~\bibnamefont{Ifju}},
  \bibinfo{journal}{The Journal of Strain Analysis for Engineering Design} \textbf{\bibinfo{volume}{36}},
  \bibinfo{pages}{101} (\bibinfo{year}{2001}).

\bibitem[{\citenamefont{Gonz{\'a}lez-Tudela and
  Cirac}(2019)}]{gonzalez2019cold}
\bibinfo{author}{\bibfnamefont{A.}~\bibnamefont{Gonz{\'a}lez-Tudela}}
  \bibnamefont{and} \bibinfo{author}{\bibfnamefont{J.~I.}~\bibnamefont{Cirac}},
  \bibinfo{journal}{Physical Review A} \textbf{\bibinfo{volume}{100}},
  \bibinfo{pages}{053604} (\bibinfo{year}{2019}).

\bibitem[{\citenamefont{O'Riordan et~al.}(2016)\citenamefont{O'Riordan, White,
  and Busch}}]{o2016moire}
\bibinfo{author}{\bibfnamefont{L.~J.} \bibnamefont{O'Riordan}},
  \bibinfo{author}{\bibfnamefont{A.~C.}~\bibnamefont{White}}, \bibnamefont{and}
  \bibinfo{author}{\bibfnamefont{T.}~\bibnamefont{Busch}},
  \bibinfo{journal}{Physical Review A} \textbf{\bibinfo{volume}{93}},
  \bibinfo{pages}{023609} (\bibinfo{year}{2016}).

\bibitem[{\citenamefont{Hu et~al.}(2020)\citenamefont{Hu, Krasnok, Mazor, Qiu,
  and Al{\`u}}}]{hu2020moire}
\bibinfo{author}{\bibfnamefont{G.}~\bibnamefont{Hu}},
  \bibinfo{author}{\bibfnamefont{A.}~\bibnamefont{Krasnok}},
  \bibinfo{author}{\bibfnamefont{Y.}~\bibnamefont{Mazor}},
  \bibinfo{author}{\bibfnamefont{C.-W.} \bibnamefont{Qiu}}, \bibnamefont{and}
  \bibinfo{author}{\bibfnamefont{A.}~\bibnamefont{Al{\`u}}},
  \bibinfo{journal}{Nano Letters} \textbf{\bibinfo{volume}{20}},
  \bibinfo{pages}{3217} (\bibinfo{year}{2020}).

\bibitem[{\citenamefont{Lu et~al.}(2021)\citenamefont{Lu, Xiaobo and Stepanov, Petr and Yang, Wei and Xie, Ming and Aamir, Mohammed Ali and Das, Ipsita and Urgell, Carles and Watanabe, Kenji and Taniguchi, Takashi and Zhang, Guangyu and others}}]{lu2019superconductors}
\bibinfo{author}{\bibfnamefont{X.}~\bibnamefont{Lu}},
  \bibinfo{author}{\bibfnamefont{P.}~\bibnamefont{Stepanov}},
  \bibinfo{author}{\bibfnamefont{W.}~\bibnamefont{Yang}},
  \bibinfo{author}{\bibfnamefont{M.} \bibnamefont{Xie}},
  \bibinfo{author}{\bibfnamefont{M.~A.} \bibnamefont{Aamir}},
  \bibinfo{author}{\bibfnamefont{I.} \bibnamefont{Das}},
  \bibinfo{author}{\bibfnamefont{C.} \bibnamefont{Urgell}},
  \bibinfo{author}{\bibfnamefont{K.} \bibnamefont{Watanabe}},
  \bibinfo{author}{\bibfnamefont{T.} \bibnamefont{Taniguchi}},
  \bibinfo{author}{\bibfnamefont{G.} \bibnamefont{Zhang}},
  \bibinfo{author}{\bibfnamefont{A.} \bibnamefont{Bachtold}},
  \bibinfo{author}{\bibfnamefont{A.~H.} \bibnamefont{MacDonald}},
  \bibnamefont{and}
  \bibinfo{author}{\bibfnamefont{D.~K.}~\bibnamefont{Efetov}},
  \bibinfo{journal}{Nature} \textbf{\bibinfo{volume}{574}},
  \bibinfo{pages}{653} (\bibinfo{year}{2019}).

\bibitem[{\citenamefont{Lou et~al.}(2021)\citenamefont{Lou, Beicheng and Zhao, Nathan and Minkov, Momchil and Guo, Cheng and Orenstein, Meir and Fan, Shanhui,}}]{lou2021theory}
\bibinfo{author}{\bibfnamefont{B.}~\bibnamefont{Lou}},
  \bibinfo{author}{\bibfnamefont{N.}~\bibnamefont{Zhao}},
  \bibinfo{author}{\bibfnamefont{M.}~\bibnamefont{Minkov}},
  \bibinfo{author}{\bibfnamefont{C.} \bibnamefont{Guo}},
  \bibinfo{author}{\bibfnamefont{M.} \bibnamefont{Orenstein}},
  \bibnamefont{and}
  \bibinfo{author}{\bibfnamefont{S.}~\bibnamefont{Fan}},
  \bibinfo{journal}{Physical Review Letters} \textbf{\bibinfo{volume}{126}},
  \bibinfo{pages}{136101} (\bibinfo{year}{2021}).

\bibitem[{\citenamefont{Sharpe et~al.}(2019)\citenamefont{Sharpe, Aaron L and Fox, Eli J and Barnard, Arthur W and Finney, Joe and Watanabe, Kenji and Taniguchi, Takashi and Kastner, MA and Goldhaber-Gordon, David}}]{sharpe2019emergent}
\bibinfo{author}{\bibfnamefont{A.~L.}~\bibnamefont{Sharpe}},
  \bibinfo{author}{\bibfnamefont{E.~J.}~\bibnamefont{Fox}},
  \bibinfo{author}{\bibfnamefont{A.~W.}~\bibnamefont{Barnard}},
  \bibinfo{author}{\bibfnamefont{J.} \bibnamefont{Finney}},
  \bibinfo{author}{\bibfnamefont{K.} \bibnamefont{Watanabe}},
  \bibinfo{author}{\bibfnamefont{T.} \bibnamefont{Taniguchi}},
  \bibinfo{author}{\bibfnamefont{M.~A.} \bibnamefont{Kastner}},
  \bibnamefont{and}
  \bibinfo{author}{\bibfnamefont{D.}~\bibnamefont{Goldhaber}},
  \bibinfo{journal}{Science} \textbf{\bibinfo{volume}{365}},
  \bibinfo{pages}{605} (\bibinfo{year}{2021}).

\bibitem[{\citenamefont{Fan et~al.}(2021)\citenamefont{Fan, Shanhui and Joannopoulos, John D,}}]{fan2002analysis}
  \bibinfo{author}{\bibfnamefont{S.}~\bibnamefont{Fan}},
  \bibnamefont{and}
  \bibinfo{author}{\bibfnamefont{J.~D.}~\bibnamefont{Joannopoulos}},
  \bibinfo{journal}{Physical Review B} \textbf{\bibinfo{volume}{65}},
  \bibinfo{pages}{235112} (\bibinfo{year}{2002}).

\bibitem[{\citenamefont{dos2007graphene.}(2016)\citenamefont{Dos Santos, JMB Lopes and Peres, NMR and Neto, AH Castro}}]{dos2007graphene}
\bibinfo{author}{\bibfnamefont{J.~M.~B.} \bibnamefont{Lopes dos Santos}},
  \bibinfo{author}{\bibfnamefont{N.~M.~R.}~\bibnamefont{Peres}}, \bibnamefont{and}
  \bibinfo{author}{\bibfnamefont{A.~H.}~\bibnamefont{Castro Neto}},
  \bibinfo{journal}{Physical Review Letters} \textbf{\bibinfo{volume}{99}},
  \bibinfo{pages}{256802} (\bibinfo{year}{2007}).

\bibitem[{\citenamefont{Sharpe et~al.}(2019)\citenamefont{Sinha, Subhajit and Adak, Pratap Chandra and Chakraborty, Atasi and Das, Kamal and Debnath, Koyendrila and Sangani, LD Varma and Watanabe, Kenji and Taniguchi, Takashi and Waghmare, Umesh V and Agarwal, Amit and others}}]{sinha2022berry}
\bibinfo{author}{\bibfnamefont{S.}~\bibnamefont{Sinha}},
  \bibinfo{author}{\bibfnamefont{P.~C.}~\bibnamefont{Adak}},
  \bibinfo{author}{\bibfnamefont{A.}~\bibnamefont{Chakraborty}},
  \bibinfo{author}{\bibfnamefont{K.} \bibnamefont{Das}},
  \bibinfo{author}{\bibfnamefont{K.} \bibnamefont{Debnath}},
  \bibinfo{author}{\bibfnamefont{L.~V.} \bibnamefont{Sangani}},
  \bibinfo{author}{\bibfnamefont{K.} \bibnamefont{Watanabe}},
  \bibinfo{author}{\bibfnamefont{T.} \bibnamefont{Taniguchi}},
  \bibinfo{author}{\bibfnamefont{U.~V.} \bibnamefont{Waghmare}},
  \bibinfo{author}{\bibfnamefont{A.} \bibnamefont{Agarwal}},
  \bibnamefont{and}
  \bibinfo{author}{\bibfnamefont{M.~M.}~\bibnamefont{Deshmukh}},
  \bibinfo{journal}{Nature Physics} \textbf{\bibinfo{volume}{18}},
  \bibinfo{pages}{765} (\bibinfo{year}{2022}).

\bibitem[{\citenamefont{Levi et~al.}(2011)\citenamefont{Cao, Yuan and Fatemi, Valla and Demir, Ahmet and Fang, Shiang and Tomarken, Spencer L and Luo, Jason Y and Sanchez-Yamagishi, Javier D and Watanabe, Kenji and Taniguchi, Takashi and Kaxiras, Efthimios and others}}]{cao2018correlated}
\bibinfo{author}{\bibfnamefont{Y.}~\bibnamefont{Cao}},
  \bibinfo{author}{\bibfnamefont{V.}~\bibnamefont{Fatemi}},
  \bibinfo{author}{\bibfnamefont{A.}~\bibnamefont{Demir}},
  \bibinfo{author}{\bibfnamefont{S.}~\bibnamefont{Fang}},
  \bibinfo{author}{\bibfnamefont{S.}~\bibnamefont{Tomarken}},
  \bibinfo{author}{\bibfnamefont{L.}~\bibnamefont{Spencer}},
  \bibinfo{author}{\bibfnamefont{J.~Y.}~\bibnamefont{Luo}},
  \bibinfo{author}{\bibfnamefont{J.~D.}~\bibnamefont{Sanchez-Yamagishi}},
  \bibinfo{author}{\bibfnamefont{K.}~\bibnamefont{Watanabe}},
  \bibinfo{author}{\bibfnamefont{T.}~\bibnamefont{Taniguchi}},
  \bibinfo{author}{\bibfnamefont{E.}~\bibnamefont{Kaxiras}},
  \bibinfo{author}{\bibfnamefont{R.~C.}~\bibnamefont{Ashoori}},
  \bibnamefont{and}
  \bibinfo{author}{\bibfnamefont{P.}~\bibnamefont{Jarillo-Herrero}},
  \bibinfo{journal}{Nature} \textbf{\bibinfo{volume}{556}},
  \bibinfo{pages}{80} (\bibinfo{year}{2018}).

\bibitem[{\citenamefont{Andrei et~al.}(2018)\citenamefont{Andrei, Eva Y and Efetov, Dmitri K and Jarillo-Herrero, Pablo and MacDonald, Allan H and Mak, Kin Fai and Senthil, T and Tutuc, Emanuel and Yazdani, Ali and Young, Andrea F}}]{andrei2021marvels}
\bibinfo{author}{\bibfnamefont{E.~Y.}~\bibnamefont{Andrei}},
  \bibinfo{author}{\bibfnamefont{D.~K.}~\bibnamefont{Efetov}},
  \bibinfo{author}{\bibfnamefont{P.}~\bibnamefont{Jarillo-Herrero}},
  \bibinfo{author}{\bibfnamefont{A.~H.}~\bibnamefont{MacDonald}},
  \bibinfo{author}{\bibfnamefont{K.~F.}~\bibnamefont{Mak}},
  \bibinfo{author}{\bibfnamefont{T.}~\bibnamefont{Senthil}}, 
  \bibinfo{author}{\bibfnamefont{E.}~\bibnamefont{Tutuc}}, \bibinfo{author}{\bibfnamefont{A.}~\bibnamefont{Yazdani}}, \bibnamefont{and}
  \bibinfo{author}{\bibfnamefont{A.~F.}~\bibnamefont{Young}},
  \bibinfo{journal}{Nature} \textbf{\bibinfo{volume}{6}}, \bibinfo{pages}{201}
  (\bibinfo{year}{2021}).

\bibitem[{\citenamefont{Du et~al.}(2018)\citenamefont{Du, Luojun and Molas, Maciej R and Huang, Zhiheng and Zhang, Guangyu and Wang, Feng and Sun, Zhipei}}]{du2023moire}
\bibinfo{author}{\bibfnamefont{L.}~\bibnamefont{Du}},
  \bibinfo{author}{\bibfnamefont{M.~R.}~\bibnamefont{Molas}},
  \bibinfo{author}{\bibfnamefont{Z.}~\bibnamefont{Huang}},
  \bibinfo{author}{\bibfnamefont{G.}~\bibnamefont{Zhang}},
  \bibinfo{author}{\bibfnamefont{F.}~\bibnamefont{Wang}},
  \bibnamefont{and}
  \bibinfo{author}{\bibfnamefont{Z.}~\bibnamefont{Sun}},
  \bibinfo{journal}{Science} \textbf{\bibinfo{volume}{379}}, \bibinfo{pages}{14}
  (\bibinfo{year}{2023}).


\bibitem[{\citenamefont{Wang et~al.}(2020)\citenamefont{Wang, Zheng, Chen,
  Huang, Kartashov, Torner, Konotop, and Ye}}]{wang2020localization}
\bibinfo{author}{\bibfnamefont{P.}~\bibnamefont{Wang}},
  \bibinfo{author}{\bibfnamefont{Y.}~\bibnamefont{Zheng}},
  \bibinfo{author}{\bibfnamefont{X.}~\bibnamefont{Chen}},
  \bibinfo{author}{\bibfnamefont{C.}~\bibnamefont{Huang}},
  \bibinfo{author}{\bibfnamefont{Y.~V.} \bibnamefont{Kartashov}},
  \bibinfo{author}{\bibfnamefont{L.}~\bibnamefont{Torner}},
  \bibinfo{author}{\bibfnamefont{V.~V.} \bibnamefont{Konotop}},
  \bibnamefont{and} \bibinfo{author}{\bibfnamefont{F.}~\bibnamefont{Ye}},
  \bibinfo{journal}{Nature} \textbf{\bibinfo{volume}{577}}, \bibinfo{pages}{42}
  (\bibinfo{year}{2020}).

\bibitem[{\citenamefont{Huang et~al.}(2016{\natexlab{a}})\citenamefont{Huang,
  Ye, Chen, Kartashov, Konotop, and Torner}}]{huang2016localization}
\bibinfo{author}{\bibfnamefont{C.}~\bibnamefont{Huang}},
  \bibinfo{author}{\bibfnamefont{F.}~\bibnamefont{Ye}},
  \bibinfo{author}{\bibfnamefont{X.}~\bibnamefont{Chen}},
  \bibinfo{author}{\bibfnamefont{Y.~V.} \bibnamefont{Kartashov}},
  \bibinfo{author}{\bibfnamefont{V.~V.} \bibnamefont{Konotop}},
  \bibnamefont{and} \bibinfo{author}{\bibfnamefont{L.}~\bibnamefont{Torner}},
  \bibinfo{journal}{Scientific Reports} \textbf{\bibinfo{volume}{6}},
  \bibinfo{pages}{1} (\bibinfo{year}{2016}{\natexlab{a}}).

\bibitem[{\citenamefont{Lahini et~al.}(2009)\citenamefont{Lahini, Pugatch,
  Pozzi, Sorel, Morandotti, Davidson, and Silberberg}}]{lahini2009observation}
\bibinfo{author}{\bibfnamefont{Y.}~\bibnamefont{Lahini}},
  \bibinfo{author}{\bibfnamefont{R.}~\bibnamefont{Pugatch}},
  \bibinfo{author}{\bibfnamefont{F.}~\bibnamefont{Pozzi}},
  \bibinfo{author}{\bibfnamefont{M.}~\bibnamefont{Sorel}},
  \bibinfo{author}{\bibfnamefont{R.}~\bibnamefont{Morandotti}},
  \bibinfo{author}{\bibfnamefont{N.}~\bibnamefont{Davidson}}, \bibnamefont{and}
  \bibinfo{author}{\bibfnamefont{Y.}~\bibnamefont{Silberberg}},
  \bibinfo{journal}{Physical Review Letters} \textbf{\bibinfo{volume}{103}},
  \bibinfo{pages}{013901} (\bibinfo{year}{2009}).

\bibitem[{\citenamefont{Billy et~al.}(2008)\citenamefont{Billy, Josse, Zuo,
  Bernard, Hambrecht, Lugan, Cl{\'e}ment, Sanchez-Palencia, Bouyer, and
  Aspect}}]{billy2008direct}
\bibinfo{author}{\bibfnamefont{J.}~\bibnamefont{Billy}},
  \bibinfo{author}{\bibfnamefont{V.}~\bibnamefont{Josse}},
  \bibinfo{author}{\bibfnamefont{Z.}~\bibnamefont{Zuo}},
  \bibinfo{author}{\bibfnamefont{A.}~\bibnamefont{Bernard}},
  \bibinfo{author}{\bibfnamefont{B.}~\bibnamefont{Hambrecht}},
  \bibinfo{author}{\bibfnamefont{P.}~\bibnamefont{Lugan}},
  \bibinfo{author}{\bibfnamefont{D.}~\bibnamefont{Cl{\'e}ment}},
  \bibinfo{author}{\bibfnamefont{L.}~\bibnamefont{Sanchez-Palencia}},
  \bibinfo{author}{\bibfnamefont{P.}~\bibnamefont{Bouyer}}, \bibnamefont{and}
  \bibinfo{author}{\bibfnamefont{A.}~\bibnamefont{Aspect}},
  \bibinfo{journal}{Nature} \textbf{\bibinfo{volume}{453}},
  \bibinfo{pages}{891} (\bibinfo{year}{2008}).

\bibitem[{\citenamefont{Goldman and Kelton}(1993)}]{goldman1993quasicrystals}
\bibinfo{author}{\bibfnamefont{A.~L.} \bibnamefont{Goldman}} \bibnamefont{and}
  \bibinfo{author}{\bibfnamefont{R.}~\bibnamefont{Kelton}},
  \bibinfo{journal}{Reviews of Modern Physics} \textbf{\bibinfo{volume}{65}},
  \bibinfo{pages}{213} (\bibinfo{year}{1993}).

\bibitem[{\citenamefont{Lifshitz and Petrich}(1997)}]{lifshitz1997theoretical}
\bibinfo{author}{\bibfnamefont{R.}~\bibnamefont{Lifshitz}} \bibnamefont{and}
  \bibinfo{author}{\bibfnamefont{D.~M.} \bibnamefont{Petrich}},
  \bibinfo{journal}{Physical Review Letters} \textbf{\bibinfo{volume}{79}},
  \bibinfo{pages}{1261} (\bibinfo{year}{1997}).

\bibitem[{\citenamefont{Efremidis et~al.}(2002)\citenamefont{Efremidis, Sears,
  Christodoulides, Fleischer, and Segev}}]{efremidis2002discrete}
\bibinfo{author}{\bibfnamefont{N.~K.} \bibnamefont{Efremidis}},
  \bibinfo{author}{\bibfnamefont{S.}~\bibnamefont{Sears}},
  \bibinfo{author}{\bibfnamefont{D.~N.} \bibnamefont{Christodoulides}},
  \bibinfo{author}{\bibfnamefont{J.~W.} \bibnamefont{Fleischer}},
  \bibnamefont{and} \bibinfo{author}{\bibfnamefont{M.}~\bibnamefont{Segev}},
  \bibinfo{journal}{Physical Review E} \textbf{\bibinfo{volume}{66}},
  \bibinfo{pages}{046602} (\bibinfo{year}{2002}).

\bibitem[{\citenamefont{Shen et~al.}(2011)\citenamefont{Shen, Tang, and
  Wang}}]{shen2011spectral}
\bibinfo{author}{\bibfnamefont{J.}~\bibnamefont{Shen}},
  \bibinfo{author}{\bibfnamefont{T.}~\bibnamefont{Tang}}, \bibnamefont{and}
  \bibinfo{author}{\bibfnamefont{L.-L.} \bibnamefont{Wang}},
  \emph{\bibinfo{title}{Spectral methods: algorithms, analysis and
  applications}}, vol.~\bibinfo{volume}{41} (\bibinfo{publisher}{Springer
  Science \& Business Media}, \bibinfo{year}{2011}).

\bibitem[{\citenamefont{Zeng et~al.}(2021)\citenamefont{Zeng, Hu, Zhang, Fu,
  Yin, Li, and Chen}}]{zeng2021localization}
\bibinfo{author}{\bibfnamefont{J.}~\bibnamefont{Zeng}},
  \bibinfo{author}{\bibfnamefont{Y.}~\bibnamefont{Hu}},
  \bibinfo{author}{\bibfnamefont{X.}~\bibnamefont{Zhang}},
  \bibinfo{author}{\bibfnamefont{S.}~\bibnamefont{Fu}},
  \bibinfo{author}{\bibfnamefont{H.}~\bibnamefont{Yin}},
  \bibinfo{author}{\bibfnamefont{Z.}~\bibnamefont{Li}}, \bibnamefont{and}
  \bibinfo{author}{\bibfnamefont{Z.}~\bibnamefont{Chen}},
  \bibinfo{journal}{Optics Express} \textbf{\bibinfo{volume}{29}},
  \bibinfo{pages}{25388} (\bibinfo{year}{2021}).

\bibitem[{\citenamefont{Davenport and
  Mahler}(1946)}]{davenport1946simultaneous}
\bibinfo{author}{\bibfnamefont{H.}~\bibnamefont{Davenport}} \bibnamefont{and}
  \bibinfo{author}{\bibfnamefont{K.}~\bibnamefont{Mahler}},
  \bibinfo{journal}{Duke Mathematical Journal} \textbf{\bibinfo{volume}{13}},
  \bibinfo{pages}{105} (\bibinfo{year}{1946}).

\bibitem[{\citenamefont{Jiang and Zhang}(2014)}]{jiang2014numerical}
\bibinfo{author}{\bibfnamefont{K.}~\bibnamefont{Jiang}} \bibnamefont{and}
  \bibinfo{author}{\bibfnamefont{P.}~\bibnamefont{Zhang}},
  \bibinfo{journal}{Journal of Computational Physics}
  \textbf{\bibinfo{volume}{256}}, \bibinfo{pages}{428} (\bibinfo{year}{2014}).

  \bibitem[{\citenamefont{Liesen}(2013)}]{liesen2013krylov}
\bibinfo{author}{\bibfnamefont{J.}~\bibnamefont{Liesen}},
\bibinfo{author}{\bibfnamefont{S.}~\bibnamefont{Zdenek}},
  \emph{\bibinfo{title}{Krylov subspace methods: principles and analysis}} (\bibinfo{publisher}{Oxford University Press}, \bibinfo{year}{2013}).

\bibitem[{\citenamefont{Joannopoulos et~al.}(2008)\citenamefont{Joannopoulos,
  Johnson, Winn, and Meade}}]{joannopoulos2008molding}
\bibinfo{author}{\bibfnamefont{J.~D.} \bibnamefont{Joannopoulos}},
  \bibinfo{author}{\bibfnamefont{S.~G.} \bibnamefont{Johnson}},
  \bibinfo{author}{\bibfnamefont{J.~N.} \bibnamefont{Winn}}, \bibnamefont{and}
  \bibinfo{author}{\bibfnamefont{R.~D.} \bibnamefont{Meade}},
  \bibinfo{journal}{Princeton Univ. Press, Princeton, NJ}
  (\bibinfo{year}{2008}).

\bibitem[{\citenamefont{Liu et~al.}(2019)\citenamefont{Liu, Sun, and
  Turner}}]{liu2019spectral}
\bibinfo{author}{\bibfnamefont{J.}~\bibnamefont{Liu}},
  \bibinfo{author}{\bibfnamefont{J.}~\bibnamefont{Sun}}, \bibnamefont{and}
  \bibinfo{author}{\bibfnamefont{T.}~\bibnamefont{Turner}},
  \bibinfo{journal}{Journal of Scientific Computing}
  \textbf{\bibinfo{volume}{79}}, \bibinfo{pages}{1814} (\bibinfo{year}{2019}).

\bibitem[{\citenamefont{Huang et~al.}(2020)\citenamefont{Huang, Sun, and
  Yang}}]{huang2020multilevel}
\bibinfo{author}{\bibfnamefont{R.}~\bibnamefont{Huang}},
  \bibinfo{author}{\bibfnamefont{J.}~\bibnamefont{Sun}}, \bibnamefont{and}
  \bibinfo{author}{\bibfnamefont{C.}~\bibnamefont{Yang}},
  \bibinfo{journal}{CSIAM Transactions on Applied Mathematics}
  \textbf{\bibinfo{volume}{1}}, \bibinfo{pages}{463} (\bibinfo{year}{2020}).

\bibitem[{\citenamefont{Kato}(2013)}]{kato2013perturbation}
\bibinfo{author}{\bibfnamefont{T.}~\bibnamefont{Kato}},
  \emph{\bibinfo{title}{Perturbation theory for linear operators}}, vol.
  \bibinfo{volume}{132} (\bibinfo{publisher}{Springer Science \& Business
  Media}, \bibinfo{year}{2013}).

\bibitem[{\citenamefont{Huang et~al.}(2016{\natexlab{b}})\citenamefont{Huang,
  Struthers, Sun, and Zhang}}]{huang2016recursive}
\bibinfo{author}{\bibfnamefont{R.}~\bibnamefont{Huang}},
  \bibinfo{author}{\bibfnamefont{A.~A.} \bibnamefont{Struthers}},
  \bibinfo{author}{\bibfnamefont{J.}~\bibnamefont{Sun}}, \bibnamefont{and}
  \bibinfo{author}{\bibfnamefont{R.}~\bibnamefont{Zhang}},
  \bibinfo{journal}{Journal of Computational Physics}
  \textbf{\bibinfo{volume}{327}}, \bibinfo{pages}{830}
  (\bibinfo{year}{2016}{\natexlab{b}}).

\bibitem[{\citenamefont{Schwartz et~al.}(2007)\citenamefont{Schwartz, Bartal,
  Fishman, and Segev}}]{schwartz2007transport}
\bibinfo{author}{\bibfnamefont{T.}~\bibnamefont{Schwartz}},
  \bibinfo{author}{\bibfnamefont{G.}~\bibnamefont{Bartal}},
  \bibinfo{author}{\bibfnamefont{S.}~\bibnamefont{Fishman}}, \bibnamefont{and}
  \bibinfo{author}{\bibfnamefont{M.}~\bibnamefont{Segev}},
  \bibinfo{journal}{Nature} \textbf{\bibinfo{volume}{446}}, \bibinfo{pages}{52}
  (\bibinfo{year}{2007}).

\bibitem[{\citenamefont{Freedman et~al.}(2006)\citenamefont{Freedman, Bartal,
  Segev, Lifshitz, Christodoulides, and Fleischer}}]{freedman2006wave}
\bibinfo{author}{\bibfnamefont{B.}~\bibnamefont{Freedman}},
  \bibinfo{author}{\bibfnamefont{G.}~\bibnamefont{Bartal}},
  \bibinfo{author}{\bibfnamefont{M.}~\bibnamefont{Segev}},
  \bibinfo{author}{\bibfnamefont{R.}~\bibnamefont{Lifshitz}},
  \bibinfo{author}{\bibfnamefont{D.~N.} \bibnamefont{Christodoulides}},
  \bibnamefont{and} \bibinfo{author}{\bibfnamefont{J.~W.}
  \bibnamefont{Fleischer}}, \bibinfo{journal}{Nature}
  \textbf{\bibinfo{volume}{440}}, \bibinfo{pages}{1166} (\bibinfo{year}{2006}).

\bibitem[{\citenamefont{Levi et~al.}(2011)\citenamefont{Levi, Rechtsman,
  Freedman, Schwartz, Manela, and Segev}}]{levi2011disorder}
\bibinfo{author}{\bibfnamefont{L.}~\bibnamefont{Levi}},
  \bibinfo{author}{\bibfnamefont{M.}~\bibnamefont{Rechtsman}},
  \bibinfo{author}{\bibfnamefont{B.}~\bibnamefont{Freedman}},
  \bibinfo{author}{\bibfnamefont{T.}~\bibnamefont{Schwartz}},
  \bibinfo{author}{\bibfnamefont{O.}~\bibnamefont{Manela}}, \bibnamefont{and}
  \bibinfo{author}{\bibfnamefont{M.}~\bibnamefont{Segev}},
  \bibinfo{journal}{Science} \textbf{\bibinfo{volume}{332}},
  \bibinfo{pages}{1541} (\bibinfo{year}{2011}).





\end{thebibliography}

\end{document}